# An $O(n^2)$ Time Algorithm for Alternating Büchi Games

Krishnendu Chatterjee[*]    Monika Henzinger[†]

**Abstract**

Computing the winning set for Büchi objectives in alternating games on graphs is a central problem in computer aided verification with a large number of applications. The long standing best known upper bound for solving the problem is $\widetilde{O}(n \cdot m)$, where $n$ is the number of vertices and $m$ is the number of edges in the graph. We are the first to break the $\widetilde{O}(n \cdot m)$ boundary by presenting a new technique that reduces the running time to $O(n^2)$. This bound also leads to $O(n^2)$ time algorithms for computing the set of almost-sure winning vertices for Büchi objectives (1) in alternating games with probabilistic transitions (improving an earlier bound of $\widetilde{O}(n \cdot m)$), (2) in concurrent graph games with constant actions (improving an earlier bound of $O(n^3)$), and (3) in Markov decision processes (improving for $m > n^{4/3}$ an earlier bound of $O(\min(m^{1.5}, m \cdot n^{2/3}))$). We also show that the same technique can be used to compute the maximal end-component decomposition of a graph in time $O(n^2)$, which is an improvement over earlier bounds for $m > n^{4/3}$. Finally, we show how to maintain the winning set for Büchi objectives in alternating games under a sequence of edge insertions or a sequence of edge deletions in $O(n)$ amortized time per operation. This is the first dynamic algorithm for this problem.

## 1 Introduction

Consider a directed graph $(V, E)$ with a partition $(V_1, V_2)$ of $V$ and a set $B \subset V$ of *Büchi* vertices. This graph is called a *game graph*. Let $n = |V|$ and $m = |E|$. Two players play the following *alternating game* on the graph that forms an infinite path. They start by placing a token on an initial vertex and then take turns indefinitely in moving the token: At a vertex $v \in V_1$ player 1 moves the token along one of the out-edges of $v$, at a vertex $u \in V_2$ player 2 moves the token along one of the out-edges of $u$. A first question to ask is given a start vertex $x \in V$ can player 1 guarantee that the infinite path visits a vertex in $B$ *at least once*, no matter what choices player 2 makes. If so player 1 can *win* from $x$ and $x$ belongs to the *winning set of player 1*. The question of computing the set of vertices from which player 1 can win (called the *winning set*) is called the *(alternating) reachability game problem*. The problem is PTIME-complete and the winning set of player 1 can be computed in time linear in the size of the graph [2, 19]. A second, more central question is whether player 1 can guarantee that the infinite path visits a vertex in $B$ *infinitely often*, no matter what choices player 2 makes. Computing the winning set of player 1 for this setting is called the *(alternating) Büchi game problem*. The best known algorithms for this problem are algorithms that repeatedly compute the alternating reachability game solution on the graph after the removal of specific vertices. Their running time is $\widetilde{O}(n \cdot m)$. We present in this paper a new algorithmic technique for the alternating Büchi game problem which is inspired by dynamic graph algorithms and which reduces the running time to $O(n^2)$.

Two-player games on graphs played by player 1 and the adversary player 2 are central in many problems in computer science, specially in verification and synthesis of systems such as the synthesis of systems from specifications and synthesis of reactive systems [11, 26, 27], verification of open systems [1], checking interface compatibility [14], well-formedness of specifications [15], and many others. Besides their application in verification, they have also been studied in artifical intelligence as AND-OR graphs [24], and in the context of alternating Turing machines [6].

---

[*]IST Austria (Institute of Science and Technology Austria) Email: `krishnendu.chatterjee@ist.ac.at`
[†]Research Group Theory and Applications of Algorithms, University of Vienna, Austria. Email: `monika.henzinger@univie.ac.at`

The class of Büchi or repeated reachability objectives was introduced in the seminal works of Büchi [3, 4, 5] in the context of automata over infinite words. The alternating Büchi game problem is one of the core problems in verification and synthesis. For example, (a) the solution of the synthesis problem for deterministic Büchi automata is achieved through solving the alternating Büchi game problem (see [23] for the importance of deterministic Büchi automata); and (b) the verification of open systems with liveness and weak fairness conditions (two key specifications used in verification) is again solved through alternating Büchi game problem [1]. Vardi [30, 29] discusses further applications of the alternating Büchi game problem and its importance. The classical algorithm for alternating Büchi games follows from the results of [16, 25, 31], its complexity is $O(n \cdot m)$. The algorithm was improved in the special case of game graphs with $m = O(n)$ to $O(n^2/\log n)$ time in [9]. A generalization of the algorithm from [9] was presented in [8], and the new algorithm requires $O((n \cdot m \cdot \log \Delta)/\log n)$ time, where $\Delta$ is the maximum out-degree. Thus the long standing best known upper bound for solving the alternating Büchi game problem is $\widetilde{O}(n \cdot m)$.

In the design and verification of open systems it is natural that the systems under verification are developed incrementally by adding choices or removing choices for the system, which is represented by player 1. However the adversary, modeled by player 2, is the environment, and the system design has no control over the environment actions. Hence there is a clear motivation to obtain dynamic algorithms for the alternating Büchi game problem, when edges leaving player-1 vertices are inserted or deleted, while edges leaving player-2 vertices remain unchanged.

**Our contributions.** In this work we present improved static and the first dynamic algorithms for the alternating Büchi game problem using graph algorithmic techniques. Our main results are as follows.
  1. We present an $O(n^2)$ time algorithm for the alternating Büchi game problem, and thus break the long standing barrier of $\widetilde{O}(n \cdot m)$ for the problem. It follows that in combination with the $O(n^2/\log n)$ algorithm for $m = O(n)$, we break the $O(n \cdot m)$ barrier for all cases.
  2. We present the first incremental and decremental algorithms for the alternating Büchi game problem for insertion and deletion of player-1 edges. Our algorithm is based on the progress measure algorithm of [20] and generalizes the Even-Shiloach algorithm for decremental reachability in undirected graphs [17]. The total time for all operations is $O(n \cdot m)$, i.e., the amortized time per operation is $O(n)$.
  3. Using our techniques to solve alternating Büchi games we also show that the maximal end-component decomposition problem (a core problem in *probabilistic* verification) can also be solved in $O(n^2)$ time (see [13] and other references of [7] for the importance of the problem). The best known bound for this problem was $O(\min(m^{1.5}, m \cdot n^{2/3}))$ [7]. Thus, our algorithm is faster for $m > n^{4/3}$ and we obtain an improved bound of $O(\min(m^{1.5}, n^2))$ for the problem.

Decremental and incremenal algorithms for computing the maximal end-component decomposition was given in [7]. However, our algorithms are the first dynamic algorithms for the alternating Büchi game problem and completely different from [7]. Our result for alternating Büchi games improves the bounds for other problems as well. We list them below.
  1. The problem of computing the set of almost-sure (or probability 1) winning vertices in alternating games with probabilistic transitions (aka simple stochastic games [12]) and Büchi objectives can be solved in $O(n^2)$ time improving the previous known $\widetilde{O}(n \cdot m)$ bound: this follows from the linear reduction of [10] from simple stochastic games to alternating Büchi games for almost-sure winning and our Büchi games algorithm.
  2. The problem of computing the set of almost-sure (probability 1) and limit-sure (probability arbitrarily close to 1) winning vertices in concurrent graph games (aka games with simultaneous interaction) with constant actions with Büchi objectives can be solved in $O(n^2)$ time: this follows from the linear reduction from concurrent games to alternating Büchi games [21] and our Büchi algorithm. The best known bound for concurrent graph games with constant actions with Büchi objectives was $O(n \cdot |\delta|)$, where $|\delta|$ is the number of transitions which is $O(n^2)$ in the worst case. Thus, in the worst case the previous best known bound was $O(n^3)$.

3. As a consequence of our $O(n^2)$ algorithm for Büchi games and the linear reduction of [10], we also obtain an $O(n^2)$ algorithm for computing almost-sure winning states for Markov decision processes with Büchi objectives. The best known bound for this problem was $O(\min(m^{1.5}, m \cdot n^{2/3}))$ [7]. Thus, our algorithm is faster for $m > n^{4/3}$ and we obtain an improved bound of $O(\min(m^{1.5}, n^2))$ for the problem.

Our main technical contribution is twofold: (1) The classical algorithm for alternating Büchi games repeatedly removes *non-winning* vertices from the game graph and then recomputes the player-1 winning set for the alternating reachability game problem. Similar to the classical algorithm our algorithm repeatedly removes non-winning vertices from the game graph. However, it finds these vertices more efficiently using a hierarchical graph decomposition technique. This technique was used first by Henzinger et al. [18] for processing repeated edge deletions in undirected graphs. We show how this technique can be extended to work for vertex deletions in (directed) game graphs. As a result we achieve faster algorithms for the alternating Büchi game problem and for computing the maximal end-component decomposition. Moreover, even in sparse graphs, our technique can be useful. If $m = c \cdot n$ and $c$ is a large constant, then our hiercharical decomposition can be used with a small number of levels, such as 2 or 3, to speed up the algorithm in practice.

(2) Even and Shiloach [17] gave a deletions-only algorithm for maintaining reachability in undirected graphs. We show how to extend this algorithm to edge deletions in directed game graphs. A purely graph-theoretic proof of the correctness of the new algorithm would be lengthy. However, by using an elegant argument based on fix-points we give a simple proof of the correctness and an analysis of the running time of the new algorithm. The new algorithm is simple and, like the algorithm in [17], does not need any sophisticated data structures. We use a "dual" fix point argument to construct an incremental algorithm for alternating Büchi games.

The paper is organized as follows: We give all necessary definitions in Section 2. Section 3 and Section 4 contain the new static algorithms for the alternating Büchi game and the maximal end-component decomposition problem. Section 5 finally contains the new dynamic algorithms.

## 2 Definitions

**Alternating Game graphs.** An *(alternating) game graph* $G = ((V, E), (V_1, V_2))$ consists of a directed graph $(V, E)$ with a set $V$ of $n$ vertices and a set $E$ of $m$ edges, and a partition $(V_1, V_2)$ of $V$ into two sets. The vertices in $V_1$ are *player 1 vertices* and the vertices in $V_2$ are *player 2 vertices*. For a vertex $u \in V$, we write $\mathsf{Out}(u) = \{v \in V \mid (u, v) \in E\}$ for the set of successor vertices of $u$ and $\mathsf{In}(u) = \{v \in V \mid (v, u) \in E\}$ for the set of incoming edges of $u$. We assume that every vertex has at least one out-going edge. i.e., $\mathsf{Out}(u)$ is non-empty for all vertices $u \in V$.

*Plays.* A game is played by two players: player 1 and player 2, who form an infinite path in the game graph by moving a token along edges. They start by placing the token on an initial vertex, and then they take moves indefinitely in the following way. If the token is on a vertex in $V_1$, then player 1 moves the token along one of the edges going out of the vertex. If the token is on a vertex in $V_2$, then player 2 does likewise. The result is an infinite path in the game graph, called *plays*. We write $\Omega$ for the set of all plays.

*Strategies.* A strategy for a player is a rule that specifies how to extend plays. Formally, a *strategy* $\sigma$ for player 1 is a function $\sigma: V_1 \to V$ such that $\sigma(v) \in \mathsf{Out}(v)$ for all $v \in V_1$, and analogously for player 2 strategies[1]. We write $\Sigma$ and $\Pi$ for the sets of all strategies for player 1 and player 2, respectively. Given a starting vertex $v \in V$, a strategy $\sigma \in \Sigma$ for player 1, and a strategy $\pi \in \Pi$ for player 2, there is a unique play, denoted $\omega(v, \sigma, \pi) = \langle v_0, v_1, v_2, \ldots \rangle$, which is defined as follows: $v_0 = v$ and for all $k \geq 0$, if $v_k \in V_1$, then $\sigma(v_k) = v_{k+1}$, and if $v_k \in V_2$, then $\pi(v_k) = v_{k+1}$.

---
[1]In general strategies are defined as functions $\sigma: V^* \cdot V_1 \to V$ that, given a finite sequence of vertices (representing the history of the play so far) which ends in a player 1 vertex, chooses the next vertex. The strategy must choose only available successors, i.e., for all $w \in V^*$ and $v \in V_1$ we have $\sigma(w \cdot v) \in \mathsf{Out}(v)$. The strategies for player 2 are defined analogously. However for all objectives considered in the paper there exists a winning

*Objectives.* We consider game graphs with a Büchi objective for player 1 and the complementary coBüchi objective for player 2. For a play $\omega = \langle v_0, v_1, v_2, \ldots \rangle \in \Omega$, we define $\mathrm{Inf}(\omega) = \{v \in V \mid v_k = v \text{ for infinitely many } k \geq 0\}$ to be the set of vertices that occur infinitely often in $\omega$. We also define reachability and safety objectives as they will be useful in the analysis of the algorithms.

1. *Reachability and safety objectives.* Given a set $T \subseteq V$ of vertices, the reachability objective $\mathrm{Reach}(T)$ requires that some vertex in $T$ be visited, and dually, the safety objective $\mathrm{Safe}(F)$ requires that only vertices in $F$ be visited. Formally, the sets of winning plays are $\mathrm{Reach}(T) = \{\langle v_0, v_1, v_2, \ldots \rangle \in \Omega \mid \exists k \geq 0.\ v_k \in T\}$ and $\mathrm{Safe}(F) = \{\langle v_0, v_1, v_2, \ldots \rangle \in \Omega \mid \forall k \geq 0.\ v_k \in F\}$. The reachability and safety objectives are dual in the sense that $\mathrm{Reach}(T) = \Omega \setminus \mathrm{Safe}(V \setminus T)$.

2. *Büchi and coBüchi objectives.* Given a set $B \subseteq V$ of vertices, the Büchi objective $\mathrm{Buchi}(B)$ requires that some vertex in $B$ be visited infinitely often, and dually, the coBüchi objective $\mathrm{coBuchi}(C)$ requires that only vertices in $C$ be visited infinitely often. Thus, the sets of winning plays are $\mathrm{Buchi}(B) = \{\omega \in \Omega \mid \mathrm{Inf}(\omega) \cap B \neq \emptyset\}$ and $\mathrm{coBuchi}(C) = \{\omega \in \Omega \mid \mathrm{Inf}(\omega) \subseteq C\}$. The Büchi and coBüchi objectives are dual in the sense that $\mathrm{Buchi}(B) = \Omega \setminus \mathrm{coBuchi}(V \setminus B)$. Observe that Büchi and coBüchi objectives are *tail (or prefix-independent)* objectives, i.e., a play satisfies the objective if and only if the play obtained by adding or deleting a finite prefix also satisfies the objective.

*Winning strategies and sets.* Given an objective $\Phi \subseteq \Omega$ for player 1, a strategy $\sigma \in \Sigma$ is a *winning strategy* for player 1 from a vertex $v$ if for all player 2 strategies $\pi \in \Pi$ the play $\omega(v, \sigma, \pi)$ is winning, i.e., $\omega(v, \sigma, \pi) \in \Phi$. The winning strategies for player 2 are defined analogously by switching the role of player 1 and player 2 in the above definition. A vertex $v \in V$ is winning for player 1 with respect to the objective $\Phi$ if player 1 has a winning strategy from $v$. Formally, the set of *winning vertices for player 1* with respect to the objective $\Phi$ is $W_1(\Phi) = \{v \in V \mid \exists \sigma \in \Sigma.\ \forall \pi \in \Pi.\ \omega(v, \sigma, \pi) \in \Phi\}$ the set of all winning vertices. Analogously, the set of all winning vertices for player 2 with respect to an objective $\Psi \subseteq \Omega$ is $W_2(\Psi) = \{v \in V \mid \exists \pi \in \Pi.\ \forall \sigma \in \Sigma.\ \omega(v, \sigma, \pi) \in \Psi\}$.

THEOREM 2.1. (CLASSICAL MEMORYLESS DETERMINACY [16]) *For all game graphs $G = ((V, E), (V_1, V_2))$, all Büchi objectives $\Phi$ for player 1, and the complementary coBüchi objective $\Psi = \Omega \setminus \Phi$ for player 2, we have $W_1(\Phi) = V \setminus W_2(\Psi)$.*

Thus the theorem shows that every vertex of $V$ either belongs to the winning set of Büchi objectives of player 1 or to the winning set of coBüchi objectives for player 2. Since we only consider this setting we simply say in the rest of the paper that every vertex either is *winning for player 1* or *winning for player 2*.

The algorithmic question in alternating graph games with Büchi objective $\Phi$ is to compute the set $W_1(\Phi)$.

## 3 Algorithms for Büchi Games

In this section we consider algorithms for Büchi games, and when we mention winning vertices or strategies we mean winning for Büchi objectives, unless explicitly mentioned otherwise. In this section we present the classical iterative algorithm for Büchi games to compute the winning sets. We then present our new algorithm. We start with the notion of *closed sets*, *attractors*, and *alternating reachability* which are key notions for the analysis of the algorithm. We present the graph theoretic definitions, and then present well-known facts that establish the connection of the graph definitions and strategies in alternating game graphs.

---

*memoryless* strategy for a player at a vertex $v$ iff there exists a winning strategy *with memory* for the player at $v$. Thus for simplicity we only consider the simpler class of memoryless strategies.

*Closed sets.* A set $U \subseteq V$ of vertices is a *closed set* for player 1 if the following two conditions hold: (a) For all vertices $u \in (U \cap V_1)$, we have $\mathsf{Out}(u) \subseteq U$, i.e., all successors of player 1 vertices in $U$ are again in $U$; and (b) for all $u \in (U \cap V_2)$, we have $\mathsf{Out}(u) \cap U \neq \emptyset$, i.e., every player 2 vertex in $U$ has a successor in $U$. The closed sets for player 2 are defined analogously as above by exchanging the roles of player 1 and player 2 (exchanging $V_1$ and $V_2$). Every closed set $U$ for player $\ell \in \{1, 2\}$, induces a sub-game graph, denoted $G \upharpoonright U$.

**Fact 1.** Consider a game graph $G$, and a closed set $U$ for player 1. Then the following assertions hold:
1. Player 2 has a winning strategy for the objective $\mathrm{Safe}(U)$ for all vertices in $U$, i.e., player 2 can ensure that if the play starts in $U$, then the play never leaves set $U$.
2. For all $T \subseteq V \setminus U$, we have $W_1(\mathrm{Reach}(T)) \cap U = \emptyset$, i.e., for any set $T$ of vertices outside $U$, player 1 does not have a strategy from vertices in $U$ to ensure to reach $T$.
3. If $U \cap B = \emptyset$ (i.e., there is no Büchi vertex in $U$), then every vertex in $U$ is winning for player 2.

*Attractors.* Given a game graph $G$, a set $U \subseteq V$ of target vertices, and a player $\ell \in \{1, 2\}$, the set $Attr_\ell(U, G)$ (called *attractor*) is the set of vertices from which player $\ell$ has a strategy to reach a vertex in $U$ against all strategies of the other player; that is, $Attr_\ell(U, G) = W_\ell(\mathrm{Reach}(U))$. The set $Attr_1(U, G)$ can be defined inductively as follows: let $R_0 = U$; let $R_{i+1} = R_i \cup \{v \in V_1 \mid \mathsf{Out}(v) \cap R_i \neq \emptyset\} \cup \{v \in V_2 \mid \mathsf{Out}(v) \subseteq R_i\}$ for all $i \geq 0$; then $Attr_1(U, G) = \bigcup_{i \geq 0} R_i$. The inductive definition of $Attr_2(U, G)$ is analogous with $V_1$ replaced by $V_2$ and vice-versa. For all vertices $v \in Attr_1(U, G)$, define $rank(v) = i$ if $v \in R_i \setminus R_{i-1}$, that is, $rank(v)$ denotes the least $i \geq 0$ such that $v$ is included in $R_i$. Define a memoryless strategy $\sigma \in \Sigma$ for player 1 as follows: for each vertex $v \in (Attr_1(U, G) \cap V_1)$ with $rank(v) = i$, choose a successor $\sigma(v) \in (R_{i-1} \cap \mathsf{Out}(v))$ (such a successor exists by the inductive definition). It follows that for all vertex $v \in Attr_1(U, G)$ and all strategies $\pi \in \Pi$ for player 2, the play $\omega(v, \sigma, \pi)$ reaches $U$ in at most $|Attr_1(U, G)|$ transitions. Observe that for $\ell \in \{1, 2\}$, we have $U \subseteq Attr_\ell(U, G)$, i.e., the set $U$ always belongs to the attractor.

*Alternating reachability.* For $\ell \in \{1, 2\}$, for a vertex $u \in Attr_\ell(U, G)$ we say that $u$ can *$alt_\ell$-reach* the set $U$. In other words, $alt_\ell$-reach denotes that player $\ell$ has a strategy to reach the target set, irrespective of the strategy of the other player.

**Fact 2.** For all game graphs $G$, all players $\ell \in \{1, 2\}$, and all sets $U \subseteq V$ of vertices, the following holds:
1. The set $V \setminus Attr_\ell(U, G)$ is a closed set for player $\ell$, i.e., no player $\ell$ vertex in $V \setminus Attr_\ell(U, G)$ has an edge to $Attr_\ell(U, G)$ and every vertex of the other player in $V \setminus Attr_\ell(U, G)$ has an edge in $V \setminus Attr_\ell(U, G)$.
2. The set $Attr_\ell(U, G)$ can be computed in time $O(|\sum_{v \in Attr_\ell(U,G)} \mathsf{In}(v)|)$ [2, 19].

COROLLARY 3.1. *Every vertex in the set $V \setminus Attr_1(B, G)$ is winning for player 2 and is not winning for player 1.*

**3.1 Classical algorithm for Büchi games** In this subsection we present the classical algorithm for Büchi games. We start with an informal description of the algorithm.

**Informal description of classical algorithm.** The *classical algorithm* (Algorithm 1) works as follows. We describe an iteration $j$ of the algorithm: the set of vertices at iteration $j$ is denoted by $V^j$, the game graph by $G^j$ and the set of Büchi vertices $B \cap V^j$ by $B^j$. At iteration $j$, the algorithm first finds the set of vertices $R^j$ from which player 1 can $alt_1$-reach the set $B^j$, i.e., computes $Attr_1(B^j, G^j)$. The rest of the vertices $Tr^j = V^j \setminus R^j$ is a closed subset for player 1, and $Tr^j \cap B^j = \emptyset$. Thus the set $Tr^j$ is winning for player 2 (by Corollary 3.1). Then the set of vertices $W_{j+1}$, from which player 2 can $alt_2$-reach the set $Tr^j$, i.e., $Attr_2(Tr^j, G^j)$ is computed. The set $W_{j+1}$ is winning for player 2, and *not* for player 1 in $G^j$ and also in $G$. Thus, it is removed from the vertex set to obtain game graph $G^{j+1}$. The algorithm then iterates on the reduced game graph, i.e., proceeds to iteration $j+1$ on $G^{j+1}$. In every iteration a linear-time attractor computation is performed with the current Büchi vertices as target to find the set of vertices which can $alt_1$-reach the Büchi set. Each iteration takes $O(m)$ time and the algorithm runs for at most $O(n)$ iterations, giving a total time of $O(n \cdot m)$. The algorithm is formally described as Algorithm 1. The correctness proof of the algorithm shows that when the algorithm terminates, all the remaining vertices are winning for player 1 [25, 28].

**Algorithm 1** Classical algorithm for Büchi Games

    **Input :** A game graph $G = ((V,E), (V_1, V_2))$ and $B \subseteq V$.
    **Output:** $W \subseteq V$.
    1. $G^0 := G$; $V^0 := V$; 2. $W_0 := \emptyset$; 3. $j := 0$
    4. **repeat**
        4.1 $W_{j+1} := \mathsf{AvoidSetClassical}(G^j, B \cap V^j)$
        4.2 $V^{j+1} := V^j \setminus W_{j+1}$; $G^{j+1} = G \upharpoonright V^{j+1}$; $j := j+1$;
    **until** $W_j = \emptyset$
    5. $W := \bigcup_{k=1}^{j} W_k$;
    6. **return** $W$.

**Procedure** AvoidSetClassical
    **Input:** Game graph $G^j$ and $B^j \subseteq V^j$.
    **Output:** set $W_{j+1} \subseteq V^j$.
    1. $R^j := Attr_1(B^j, G^j)$; 2. $Tr^j := V^j \setminus R^j$; 3. $W_{j+1} := Attr_2(Tr^j, G^j)$

THEOREM 3.1. (CORRECTNESS AND RUNNING TIME) *Given a game graph $G = ((V, E), (V_1, V_2))$ and $B \subseteq V$ the following assertions hold:*
1. *$W = W_2(\text{coBuchi}(V \setminus B))$ and $V \setminus W = W_1(\text{Buchi}(B))$, where $W$ is the output of Algorithm 1; and*
2. *the running time of Algorithm 1 is $O(n \cdot m)$.*

**3.2 New Algorithm** In this section we present our new algorithm for computing the winning set for game graphs with Büchi objectives in time $O(n^2)$.

**Notations.** Given an alternating game graph $G = ((V, E), (V_1, V_2))$ and a set $B$ of Büchi vertices, we label the Büchi vertices as priority 0 vertices, and the set $V \setminus B$ as priority 1 vertices. For every vertex $v$ the inedges have a *fixed* order such that all edges from priority 1 player-2 vertices come before all other edges. We maintain $\log n$ graphs $G_i$ such that $G_i = (V, E_i)$. The set $E_i$ contains all edges $(u, v)$ where (a) $\mathsf{outdeg}(u) \le 2^i$, where $\mathsf{outdeg}(u) = |\mathsf{Out}(u)|$ or (b) the edge $(u,v)$ belongs to the first $2^i$ inedges of vertex $v$. Note that $E_{i-1} \subseteq E_i$ since the order of the inedges is fixed. We color every player-1 vertex $v$ in $G_i$ *blue* if $\mathsf{outdeg}(v) > 2^i$. We color every player-2 vertex $v$ in $G_i$ *orange* if $\mathsf{outdeg}(v) > 2^i$. All other vertices have color white. For every vertex $v$ that is white in $G_i$, all its outedges $\mathsf{Out}(v)$ are contained in $E_i$. These edges add up to $2^i \cdot n$ edges to $E_i$. Additionally the first up to $2^i$ inedges of every vertex belong to $E_i$, adding another up to $2^i \cdot n$ edges to $E_i$. Thus $|E_i| \le 2^{i+1} \cdot n$. We denote by $G$ the full graph. Note that $G = G_{\log n}$ and thus all vertices in $G_{\log n}$ are white.

**The new algorithm** NEWALGO. The new algorithm consists of two nested loops, an outer loop with loop counter $j$ and an inner loop with loop counter $i$. The algorithm will iteratively delete vertices from the graph, and we denote by $D_j$ the set of vertices deleted in iteration $j$, and by $U$ the set of vertices deleted in all iterations upto the current iteration (initially $U$ is empty). For $j \ge 1$, we will denote by $G_i^j$ the sub-graph of $G_i$ induced after removal of the set $U$ of vertices at the beginning of iteration $j$, and $G_i^0$ is $G_i$ (the initial graphs). We denote the vertex set in iteration $j$ as $V^j$ and the Büchi set as $B^j$ (i.e., $B^j := V^j \cap B$). The intuitive description of the algorithm is as follows: Starting from $i = 0$ the algorithm searches in each iteration $j$ in each graph $G_i^j$ for a special player-1 closed set $S_j$ with no Büchi vertex and stop at the smallest $i$ at which such a closed set exists. Since $S_j \cap B^j = \emptyset$, Fact 1 implies that all the vertices in $S_j$ are winning for player 2. Thus, by the same arguments as for the classical algorithm the player-2 attractor $Attr_2(S^j, G_i^j)$ are winning for player 2 in $G_i^j$ and, as our correctness proof shows, also winning in $G$. Thus they are removed from the vertex set and the algorithm iterates on the reduced game graph. Computing $S_j$ takes time $O(2^i \cdot n)$ and, due to the fact that no such set was found in $G_{i-1}^j$ we can show that $S_j$ it contains at least $2^{i-1}$ vertices. Thus, using amortized analysis we charge $O(n)$ to each of the $2^{i-1}$ vertices in $S_j$ that are removed, giving a total running time of $O(n^2)$. The details of NEWALGO follow.

1. For $j = 0$, let $Y_0 := Attr_1(B, G^0)$ (where $G^0$ is the initial game graph); $X_0 := V \setminus Y_0$ (i.e., $X_0$ is the set of vertices that cannot alt$_1$-reach the Büchi vertices in the initial game graph $G$); and compute $D_0 := Attr_2(X, G)$ using attractor computation.
2. Remove the vertices of $D_j$ from all $\log n$ graphs $G_i^j$ to create graphs $G_i^{j+1}$; $j := j + 1$; and $U := U \cup D_j$;
3. $i := 1$;
4. repeat
   (a) Let $Z_i^j$ be the vertices of $V^j$ that are (i) either orange *with no outedges* in $G_i^j$ or (ii) blue in $G_i^j$.
   (b) Compute the set $Y_i^j$ of vertices in $G_i^j$ that can alt$_1$-reach the Büchi vertices or $Z_j^i$, i.e., compute $Y_i^j := Attr_1(B^j \cup Z_i^j, G_i^j)$ using attractor computation.
   (c) $S_j := V^j \setminus Y_i^j$ (i.e., $V^j \setminus Attr_1(B^j \cup Z_i^j, G_i^j)$); $i := i + 1$
5. until $S_j$ is non-empty or $i = \log n$
6. if $S_j \neq \emptyset$, then $D_j := Attr_2(S_j, G^j)$ and go to Step 2, else the whole algorithm terminates and outputs $V \setminus U$.

Let $U^*$ be the set of vertices removed from the graph over all iterations and $Y^* = V \setminus U^*$ be the output of the algorithm. We first show that $Y^* \subseteq W_1(\Phi)$, where $\Phi$ is the Büchi objective, i.e., $Y^*$ is winning for player 1. Then we show that $U^* \cap W_1(\Phi) = \emptyset$ (i.e., $U^*$ is not winning for player 1). Together with Theorem 2.1 this shows that $Y^* = W_1(\Phi)$ establishing the correctness of the algorithm. Finally we analyze the running time of the algorithm.

LEMMA 3.1. *Let $Y^*$ be the output of NEWALGO, and let $G^*$ and $B^*$ be the game graph and the Büchi set on termination, respectively (i.e., $G^*$ is the graph induced by $Y^*$ and $B^*$ is $B \cap Y^*$). The following assertions hold:*
1. $Y^* = Attr_1(B^*, G^*)$, *i.e., player 1 can alt$_1$-reach the set $B^*$ in $G^*$ from $Y^*$.*
2. $Y^*$ *is a player-2 closed set in the original game graph $G$.*
3. $Y^* \subseteq W_1(\Phi)$, *where $\Phi$ is the Büchi objective.*

*Proof.* We prove the three parts below.
1. Consider the last iteration $j^*$ of the outer loop of the algorithm. Since it is the last iteration, the set $S_{j^*}$ must be empty. It follows that $i$ must have been $\log n$ in the last iteration of the repeat loop, i.e., the last iteration of the repeat loop considered $G_{\log n}^{j^*} = G^*$. Let $i = \log n$. Note that all vertices are white in $G^*$, i.e., $Z_i^{j^*}$ was empty. Hence we have $Y_i^{j^*} = Attr_1(B^* \cup Z_i^{j^*}, G^*) = Attr_1(B^*, G^*)$. Hence the fact that $S_{j^*}$ was empty at the end of the iteration implies that $V^{j^*} \setminus Y_i^{j^*}$ was empty, i.e., that all vertices of $G^*$ belong to $Attr_1(B^*, G^*)$. Hence $Y^* = Attr_1(B^*, G^*)$.
2. Whenever a set of vertices is deleted in any iteration, it is an player-2 attractor. Hence if a vertex $u \in Y^* \cap V_2$ would have an edge to a vertex $v \in U^*$, then $u$ would have been included in $U^*$ (where $U^* = V \setminus Y^*$). Similarly for a player 1 vertex $u \in Y^* \cap V_1$ it must have an edge in $Y^*$, as we assume that it has at least one out-edge and if all its out-edges pointed to $U^*$ it would have been included in $U^*$. It follows that $Y^*$ is a player-2 closed set in $G$.
3. The result is obtained from the previous two items. Consider a memoryless attractor strategy $\sigma$ in $G^*$ for player-1 that ensures that for all vertices in $Y^*$ the set $B^*$ is reached within $|Y^*|$ steps against all strategies of player-2. Moreover the strategy only chooses successor in $Y^*$. Since $Y^*$ is a player-2 closed set, it follows that against all strategies of player-2 the set $Y^*$ is never left, thus it is ensured that $B^*$ is visited infinitely often. Hence the strategy $\sigma$ ensures that for all vertices $v \in Y^*$ and all strategies $\pi$ we have $\omega(v, \sigma, \pi) \in \Phi$. It follows that $Y^* \subseteq W_1(\Phi)$.

The desired result follows. ∎

To complete the correctness proof we need to show that if $U^* = V \setminus Y^*$, then $U^* \cap W_1(\Phi) = \emptyset$, where $\Phi$ is the Büchi objective. We will show the result by induction on the number of iterations. Let us denote by $U_j$ the set of vertices removed till iteration $j$. The base case is trivial as initially $U$ is emptyset. By inductive hypothesis, we assume for $j \geq 1$ we have $U_{j-1} \cap W_1(\Phi) = \emptyset$, and then show that $U_j \cap W_1(\Phi) = \emptyset$. Let $G^j$ be the alternating game graph obtained after removal of the set $U_{j-1}$ of vertices. We will show the following claim.

**Claim 1.** In $G^j$, let $S_j$ be the non-empty set identified in iteration $j$, then $Attr_1(B^j, G^j) \cap S_j = \emptyset$.

In the following two lemmata we first show how with Claim 1 we establish the correctness of our algorithm and finally prove Claim 1 to complete the correctness proof.

LEMMA 3.2. *The inductive hypothesis that $U_{j-1} \cap W_1(\Phi) = \emptyset$ and Claim 1 implies that $S_j \cap W_1(\Phi) = \emptyset$.*

*Proof.* By Claim 1 we have $Attr_1(B^j, G^j) \cap S_j = \emptyset$, and it follows that if player 1 follows a strategy from any vertex in $S_j$ such that the set $V^j = V \setminus U_{j-1}$ of vertices is never left, then no Büchi vertex is ever reached. If the set $V^j$ is left after a finite number of steps, then the set $U_{j-1}$ is reached, and by inductive hypothesis $U_{j-1} \cap W_1(\Phi) = \emptyset$, i.e., player 2 can ensure from $U_{j-1}$ that the set of Büchi vertices is visited finitely often. Since the Büchi objective is independent of finite prefixes, it follows that if $V^j$ is left and $U_{j-1}$ is reached, then player 2 ensures that the Büchi objective is not satisfied. It follows that $S_j \cap W_1(\Phi) = \emptyset$. ∎

LEMMA 3.3. *The inductive hypothesis that $U_{j-1} \cap W_1(\Phi) = \emptyset$ and Claim 1 implies that $U_j \cap W_1(\Phi) = \emptyset$.*

*Proof.* Observe that $U_j \setminus U_{j-1}$ is obtained as a player 2 attractor to $S_j$, and hence player 2 can ensure from $U_j \setminus U_{j-1}$ that $S_j$ is reached in finite number of steps. Since Büchi objective is independent of finite prefixes, by inductive hypothesis $U_{j-1} \cap W_1(\Phi) = \emptyset$, and by Lemma 3.2 we have that $S_j \cap W_1(\Phi) = \emptyset$, it follows that $U_j \cap W_1(\Phi) = \emptyset$. ∎

Hence to complete the proof we need to establish Claim 1 and this is achieved in the following two lemmata. We start with the notion of a separating cut.

**Separating cut.** We say a set $S$ of vertices induces a *separating cut* in a graph $G_i$ or $G_i^j$ if (a) the only edges from $S$ to $V \setminus S$ come from player-2 vertices in $S$, (b) every player-2 vertex in $S$ has an edge to another vertex in $S$, (c) every player-1 vertex in $S$ is white, and (d) $B \cap S = \emptyset$. Thus $S$ is a player-1 closed set where every player-1 vertex is white and which does not contain a vertex in $B$.

LEMMA 3.4. *Let $G = ((V, E), (V_1, V_2))$ be a game graph where every vertex has at least outdegree 1, and $G' = ((V, E'), (V_1, V_2))$ be a sub-graph of $G$ with $E' \subseteq E$. Let $Z$ be a set of blue player-1 and orange player-2 vertices of $G'$ such that all orange vertices have outdegree 0 in $G'$. If $S$ induces a separating cut in $G'$, then no vertex of $S$ belongs to $Attr_1(B \cup Z, G)$.*

*Proof.* We first show that every vertex in $S$ has an edge to another vertex in $S$ in $G'$. For player-2 vertices this follows from condition (b) of a separating cut. For player-1 vertices this follows since they have outdegree 1 in $G$, are white in $G'$, and cannot have an edge to a vertex in $V \setminus S$.

Note that $S \cap (B \cup Z) = \emptyset$ since $S$ contains no blue vertex of $G_i$, every orange vertex in $S$ has outdegree at least 1 and $B \cap S = \emptyset$ by condition (d) of a separating cut. By condition (a) for all player-1 vertices in $S$ all out-going edges are in $S$. It follows that $S$ is a player-1 closed set, and since $S \cap (B \cup Z) = \emptyset$, the result follows from Fact 1. ∎

LEMMA 3.5. *We have $S_j \cap Attr_1(B^j, G^j) = \emptyset$.*

*Proof.* Let $v$ be a vertex in $S_j$. By construction $v$ cannot alt$_1$-reach $B^j \cup Z_{i^*}^j$ in $G_{i^*}^j$, where $i^*$ was the last value of $i$ in the repeat loop of iteration $j$. We will show that $v$ cannot alt$_1$-reach $B^j$ in $G^j$. If suffices to show that $S_j$ induces a separating cut in $G^j$. Then we can simply apply Lemma 3.4 with $G = G^j$, $G' = G_{i^*}^j$, $Z = \emptyset$, and $S = S_j$ to prove the lemma.

1. *Condition (a).* By construction no player-1 vertex in $S_j$ has an edge to $V^j \setminus S_j$, otherwise it would belong to the player-1 attractor of $B^j \cup Z^j_{i*}$. Since all player-1 vertices in $S_j$ are white in $G^j_{i*}$, the outedges of the player-1 vertices in $S_j$ are the same in $E^j_{i*}$ and in $E^j$. Thus condition (a) of a separating cut holds in $G^j$.
2. *Condition (b).* Every player-2 vertex in $S_j$ must have an edge to another vertex in $S_j$, otherwise all its edges would go to vertices in $V^j \setminus S_j$ and thus it would belong to $Attr_1(B^j \cup Z^j_{i*}, G^j_{i*})$. Since $E^j_{i*} \subseteq E^j$, the same holds in $G^j$. Hence condition (b) of a separating cut holds in $G^j$.
3. *Condition (c).* All vertices are white in $G^j$. Thus condition (c) holds trivially.
4. *Condition (d).* The condition (d), $S_j \cap B^j = \emptyset$ holds, since otherwise a vertex of $S_j$ would belong to $B^j$ and, thus, to $Attr_1(B^j \cup Z^j_{i*}, G^j_{i*})$.

Thus $S_j$ induces a separating cut in $G^j$. The desired result follows. ∎

Lemma 3.5 proves Claim 1 and this completes the correctness proof, and gives the following lemma.

LEMMA 3.6. *Let $Y^*$ be the output of* NEWALGO. *Then we have $Y^* = W_1(\Phi)$, where $\Phi$ is the Büchi objective.*

**Running time analysis.** We now analyze the running time of the algorithm.

LEMMA 3.7. *Let $G^j_i$ be a game graph in iteration $j$ and let $Z^j_i$ be the set of blue and degree-0 orange vertices of $G^j_i$ as defined in iteration $j$ of the outer loop and $i$ of the inner loop of the algorithm. If $S$ induces a separating cut in $G^j_i$, then $S \subseteq S_j$.*

*Proof.* None of the vertices in $S$ can $alt_1$-reach $B$ in $G^j$ by Lemma 3.5. By Lemma 3.4 none of the vertices in $S$ can $alt_1$-reach $B^j \cup Z^j_i$. Hence we have $S \subseteq V^j \setminus Attr_1(B^j \cup Z^j_i, G^j_i)$. Thus $S \subseteq S_j$. ∎

Since $S_j$ is a complement of a player-2 attractor it is a player-1 closed set, all vertices in $S_j$ are white, and there is no Büchi vertex in $S_j$. Hence $S_j$ is a separating cut. The previous lemma shows that every separating cut $S$ is a subset of $S_j$. It follows that $S_j$ is the largest (under set inclusion) separating cut.

LEMMA 3.8. *The total time spent by* NEWALGO *is $O(n^2)$.*

*Proof.* We present the $O(n^2)$ running time analysis and we consider two cases.

*All other than the last iteration of the outer loop.* Assume in iteration $j$ the algorithm stops the repeat until loop at value $i$ and this is not the last iteration of the algorithm. Then $S_j$ is not empty. Note that all player-1 vertices in $S_j$ are white, since $Z^j_i$ contains all blue player-1 vertices of $V^j$ and $S_j = V^j \setminus Attr_1(B^j \cup Z^j_i, G^j_i)$. Thus, $S_j$ induces a separating cut in $G^j_i$. Consider the set $S_j$ in $G^j_{i-1}$. There are 2 cases to consider:

*Case 1:* $S_j$ contains a player-1 vertex $x$ that is blue in $G^j_{i-1}$. Thus $x$ has outdegree at least $2^{i-1}$ in $G^j_i$ and none of these edges go to vertices in $V^j \setminus S_j$ in $G^j_i$. Thus, $S_j$ contains at least $2^{i-1}$ vertices.

*Case 2:* All player-1 vertices in $S_j$ are white in $G^j_{i-1}$. Thus, their outedges in $G^j_i$ and $G^j_{i-1}$ are identical.

Note that no priority-1 player-2 vertices in $V^j \setminus S_j$ point to vertices of $S_j$ in $G^j_i$. Since $E^j_{i-1} \subseteq E^j_i$ it follows that no priority-1 player-2 vertices in $V^j \setminus S_j$ point to vertices in $S_j$ in $G^j_{i-1}$. Consider a player-2 vertex $u$ in $S_j$. Thus there exists an edge $(u,v) \in E^j_i$ with $v \in S_j$. There are two possibilities.

*Case 2a:* For all player-2 vertices $u \in S_j$ there exists a vertex $v \in S_j$ with $(u,v) \in E^j_{i-1}$. But then $S_j$ would be a separating cut in $G^j_{i-1}$. By Lemma 3.7 it follows that $S_j$ would be non-empty in iteration $i-1$ and thus the repeat loop would have stopped after iteration $i-1$. This is not the case and thus the condition of Case 2a does not hold.

*Case 2b:* There exists a player 2 vertex $u \in S_j$ that has an edge $(u,v) \in E^j_i$ to a vertex $v \in S_j$ but this edge is not contained in $E^j_{i-1}$. This can only happen if $u$ is orange in $G^j_{i-1}$ and $v$ has $2^{i-1}$ other inedges in $E^j_{i-1}$. Since the edge $(u,v)$ where $u$ is a priority-1 player-2 vertex is not in $G^j_{i-1}$, all inedges of $v$ that are in $G^j_{i-1}$ are from priority-1 player-2 vertices by the fixed order of inedges. It follows that none of the inedges of $v$ in $G_{i-1}^j$ are from $V^j \setminus S_j$ and, thus, $S_j$ must contain at least $2^{i-1}$ player-2 vertices.

Thus in either case $S_j$ contains at least $2^{i-1}$ vertices and all these vertices are deleted. The time spent for all the executions of the repeat loop in this iteration of the outer loop it the time spent in all graphs $G_1, G_2, ..., G_{i^*}$, which sums to $O(2^i \cdot n)$. We charge $O(n)$ work to each deleted vertex. This accounts for all but the last iteration of the outer loop. As the algorithm deletes at most $n$ vertices the total time spent over the whole algorithm other than the last iteration is $O(n^2)$.

*The last iteration of the outer loop.* In the last iteration of the outer loop, when no vertex is deleted, the algorithm works on all $\log n$ graphs, spending time $O(n \cdot 2^i)$ in graph $G_i$. Since there are $\log n$ graphs, the total time is $O(n \cdot 2 \cdot 2^{\log n}) = O(n^2)$. An identical argument also shows that the time to built all the initial graphs $G_i$ is at most $O(n^2)$. Hence the desired result follows. ∎

THEOREM 3.2. *Given a game graph $G$ with $n$ vertices, and an Büchi objective $\Phi$, algorithm* NEWALGO *correctly computes the winning set $W_1(\Phi)$ in time $O(n^2)$.*

## 4 Maximal End-component Decomposition Algorithm

In this section we present an algorithm for the maximal end-component decomposition problem that runs in $O(n^2)$ time. The maximal end-component problem is the core algorithmic problem in verification of probabilistic systems, and the graph theoretic description of the problem for game graphs is defined below.

**Maximal end-component decomposition.** Given a game graph $G = ((V, E), (V_1, V_2))$, an *end-component* $U \subseteq V$ is a set of vertices such that (a) the graph $(U, E \cap U \times U)$ is strongly connected; (b) for all $u \in U \cap V_2$ and all $(u, v) \in E$ we have $v \in U$; and (c) either $|U| \geq 2$, or $U = \{v\}$ and there is a self-loop at $v$ (i.e., $(v, v) \in E$). In other words, an end-component is a player-2 closed set that is strongly connected. Note that if $U_1$ and $U_2$ are end-components with $U_1 \cap U_2 \neq \emptyset$, then $U_1 \cup U_2$ is an end-component. A *maximal end-component (mec)* is an end-component that is maximal under set inclusion. Every vertex of $V$ belongs to *at most* one maximal end-component. The *maximal end-component (mec) decomposition* consists of all the maximal end-components of $V$ and all vertices of $V$ that do not belong to *any* maximal end-component. Maximal end-components generalize strongly connected components (scc's) for directed graphs (with $V_2 = \emptyset$) and closed recurrent sets for Markov chains (with $V_1 = \emptyset$).

**Notations.** Given a game graph $G = ((V, E), (V_1, V_2))$, we will denote by $Reachable(X, G)$ the set of vertices that can reach a vertex in $X$ in the graph $(V, E)$. Note that $X \subseteq Reachable(X, G)$. We maintain $\log n$ graphs $G_i$ such that $G_i = (V, E_i)$ and $E_i$ contains all edges $(u, v)$ where $\mathsf{o}utdeg(u) \leq 2^i$. We denote by $G$ the full graph. We color vertices $v$ in $G_i$ *blue* if $\mathsf{o}utdeg(v) > 2^i$, i.e., $\mathsf{Bl}_i = \{v \in V \mid \mathsf{o}utdeg(v) > 2^i\}$ and all other vertices are colored *white*, i.e., $\mathsf{Wh}_i = \{v \in V \mid \mathsf{o}utdeg(v) \leq 2^i\}$. Note that $G = G_{\log n}$ and thus all vertices in $G_{\log n}$ are white. Thus, none of the outedges of the blue vertices of $G_i$ belong to $G_i$, i.e., all blue vertices have outdegree 0 in $G_i$. A *bottom* scc $C$ of a graph is a scc that has no edge leaving out of $C$. Every graph has a bottom scc and every bottom scc is a mec.

**Maximal end-component decomposition algorithm.** The algorithm consists of two nested loops, an outer loop with loop counter $j$ and an inner loop with loop counter $i$. The algorithm will iteratively delete vertices from the graph, and we denote by $D_j$ the set of vertices deleted in iteration $j$. We will denote by $G_i^j$ the sub-graph of $G_i$ at the beginning of iteration $j$ (as for NEWALGO) and the vertex set in iteration $j$ is denoted as $V^j$. The set $\mathsf{Bl}_i^j$ is the set of vertices in $G_i^j$ with outdegree greater than $2^i$ in $G_i^j$. Basically the algorithm is similar to NEWALGO, and instead of searching for separating cuts, the algorithm for mec decomposition searches for bottom scc's. The steps of the algorithm are as follows and we refer the algorithm as NEWMECALGO.

1. Let $D_j$ be the set of vertices deleted in iteration $j$. For $j := 0$, let $D_0 := Attr_2(X, G^0)$, where $X$ is the set of vertices that are in the bottom scc's in the initial graph $G$. Every bottom scc is an mec and included in the mec decomposition.

2. Remove the vertices of $D_j$ from all $\log n$ graphs $G_i^j$ to create graph $G_i^{j+1}$; $j := j + 1$. If all vertices are removed, then the whole algorithm terminates and outputs the mec decomposition.
3. $i := 1$;
4. repeat
   (a) Compute all the vertices in $G_i^j$ that can reach the blue vertices using the standard linear-time algorithm for reachability.
   (b) Let $S_j = V^j \setminus Reachable(\mathsf{Bl}_i^j, G_i^j)$ be the set of vertices that cannot reach the set $\mathsf{Bl}_i^j$ blue vertices in $G_i^j$; $i := i + 1$
5. until $S_j$ is non-empty
6. if $S_j \neq \emptyset$, then let $D_j := Attr_2(X, G^j)$, where $X$ is the set of vertices that are in the bottom scc's in the sub-graph induced by $S_j$ in $G_i^j$. Every bottom scc is an mec and included in the mec decomposition. Go to Step 2.

**Basic correctness argument.** Let us denote $G^j$ to be the remaining game graph after iteration $j$. Let $S_j$ be the set identified at iteration $j$, and let the inner iteration stop at $i^*$. All vertices in $S_j$ are white, since $S_j = V^j \setminus Reachable(\mathsf{Bl}_{i^*}^j, G_{i^*}^j)$ and $\mathsf{Bl}_{i^*}^j \subseteq Reachable(\mathsf{Bl}_{i^*}^j, G_{i^*}^j)$. For all $v \in S_j$, all outedges from $v$ end in a vertex in $S_j$: otherwise if there is an edge from $v$ to $Reachable(\mathsf{Bl}_{i^*}^j, G_{i^*}^j)$, then $v$ would have been included in $Reachable(\mathsf{Bl}_{i^*}^j, G_{i^*}^j)$. Hence any bottom scc in the subgraph induced by $S_j$ in $G_{i^*}^j$ is also a bottom scc of $G^j$. The correctness of the identification the bottom scc as an mec and removal of the attractor follows from the following two lemmata established in [7] (see Lemma 2.1 and Lemma 2.2 of [7]). The first lemma below establishes that the player-2 attractor of a mec and the player-2 attractor of certain vertices of an scc do not belong to any mec and that it, thus, can be removed without affecting the mec decomposition of the remaining graph. Hence, the lemma is used to identify vertices that do not belong to *any* mec. The second lemma below shows under which condition an scc is an mec. Thus, it can be used to identify vertices that *form* a mec. It follows trivially from the second lemma that every bottom scc is a mec.

LEMMA 4.1. ([7]) *Let $G = ((V, E), (V_1, V_2))$ be a game graph, and let $(V, E)$ be the graph.*

1. *Let $C$ be a scc in $(V, E)$. Let $U = \{v \in C \cap V_2 \mid \mathsf{Out}(v) \cap (V \setminus C) \neq \emptyset\}$ be the player-2 vertices in $C$ with edges out of $C$. Let $Z = Attr_2(U, G) \cap C$. Then for all non-trivial mec's $X$ in $G$ we have $Z \cap X = \emptyset$ and for any edge $(u, v)$ with $u \in X$ and $v \in Z$, $u$ must belong to $V_1$.*

2. *Let $C$ be a mec in $G$. Let $Z = Attr_2(C, G) \setminus C$. Then for all non-trivial mec's $X$ with $X \neq C$ in $G$ we have $Z \cap X = \emptyset$ and for any edge $(u, v)$ with $u \in X$ and $v \in Z$, $u$ must belong to $V_1$.*

LEMMA 4.2. ([7]) *Let $G = ((V, E), (V_1, V_2))$ be a game graph, and let $(V, E)$ be the graph. Let $C$ be a scc in $(V, E)$ such that for all $v \in C \cap V_2$ we have $\mathsf{Out}(v) \subseteq C$. Then $C$ is a mec.*

The correctness of the algorithm follows.

LEMMA 4.3. *Algorithm NEWMECALGO correctly computes the mec decomposition of a game graph.*

**Running time analysis.** The crucial result of the running time analysis depends on the following lemma. It shows that in an outer iteration $j$, if the inner iteration stops at iteration $i^*$ and $X$ is the set of vertices identified as bottom scc, then $X \cap \mathsf{Bl}_{i^*-1}^j$ is non-empty.

LEMMA 4.4. *Consider an outer iteration $j$ of the algorithm, and let the inner iteration stop at iteration $i^*$. Let $X$ be hte set of vertices identified as bottom scc of the graph induced by $S$ in $G_{i^*}^j$. Then $X \cap \mathsf{Bl}_{i^*-1}^j \neq \emptyset$.*

*Proof.* Assume towards contradiction that there is a bottom scc $C$ in the induced subgraph of $S$ in $G_{i^*}^j$ such that $C \cap \mathsf{Bl}_{i^*-1}^j = \emptyset$. Now we consider the iteration $i^* - 1$ and then for every vertex in $C$ in $G_{i^*-1}^j$ all outedges end in a vertex in $C$. Since $C$ does not contain a vertex from $\mathsf{Bl}_{i^*-1}^j$ and $C$ has no outgoing edges, it follows that $C \subseteq V^j \setminus \mathit{Reachable}(\mathsf{Bl}_{i^*-1}^j, G_{i^*}^j)$. Since all edges of $G_{i^*-1}^j$ are contained in $G_{i^*}^j$ it follows that $C \subseteq V^j \setminus \mathit{Reachable}(\mathsf{Bl}_{i^*-1}^j, G_{i^*-1}^j)$. It follows that a non-emptyset $S_j$ would have been identified in iteration $i^* - 1$, and this contradicts that the algorithm stops at iteration $i^*$ and not in $i^* - 1$. ∎

LEMMA 4.5. *The total time spent by* NEWMECALGO *is* $O(n^2)$.

*Proof.* Assume that for an outer iteration $j$, the inner iteration stops the repeat until loop at value $i^*$. By the previous lemma, one of the vertices $v$ in $X$ must have belong to $\mathsf{Bl}_{i^*-1}^j$ and thus it has outdegree at least $2^{i^*-1}$. Since we identify the bottom scc that contain $v$ it must contain all the endpoints of the outedges from $v$. Hence $X$ contains at least $2^{i^*-1}$ vertices. The time spent for all the executions of the repeat loop in this iteration of the outer loop it the time spent in all graphs $G_1^j, G_2^j, ..., G_{i^*}^j$, which sums to $O(2^{i^*} \cdot n)$. We charge $O(n)$ to each deleted vertex. As the algorithm deletes at most $n$ vertices the total time spent over the whole algorithm is $O(n^2)$. The removal of all the player-2 attractors overall iterations takes $O(m) = O(n^2)$ time. Similar to the proof of Lemma 3.8, the time required to built all the initial graphs $G_i$ is at most $O(n^2)$. The result follows. ∎

THEOREM 4.1. *Algorithm* NEWMECALGO *correctly computes the mec decomposition of a game graph in* $O(n^2)$ *time.*

## 5 Decremental and Incremental Algorithms

In this section we present the decremental and incremental algorithms for computing the winning set in game graphs with Büchi objectives. We will show that the *progress measure* algorithm of [20] works in total time $O(n \cdot m)$ for a sequence of player-1 edge deletions (or insertions), and hence the amortized time per operation is $O(n)$. Since Büchi objectives generalize reachability objectives, and alternating game graphs generalize directed graphs, our algorithm is a generalization of the Even-Shiloach algorithm [17] for decremental reachability in graphs. However our proof is very different, based on a fix-point argument, and is much simpler. We first present the algorithm for the decremental case.

**5.1 Decremental algorithm for Büchi games** In this section we present the decremental algorithm, and we consider only deletion of player-1 edges. Our decremental algorithm is based on the notion of progress measure and we start with the notion of a progress measure and valid progress measure.

*Progress measure.* Given a game graph with $n$ vertices, a progress measure is a function $\rho : V \to [n] \cup \top$, where $[n] = \{0, 1, 2, \ldots, n\}$, that assigns to every vertex either a number from 0 to $n$, or the top element $\top$. We will follow the conventions that: (a) for all $j \in [n]$ we have $j < \top$; (b) $n + 1 = \top$; (c) $\top + 1 = \top$; (d) $\top \geq \top$. Given a game graph with a set $B$ of Büchi vertices, a progress measure $\rho$ is a *valid* progress measure if the following conditions hold for all $v \in V$: (i) for $v \in V_1 \cap B$, we have $\rho(v) = \top$ if for all $(v, w) \in E$ we have $\rho(w) = \top$, and 0 otherwise; (ii) for $v \in V_2 \cap B$, we have $\rho(v) = \top$ if there exists $(v, w) \in E$ with $\rho(w) = \top$, and 0 otherwise; (iii) for $v \in V_1 \setminus B$, we have $\rho(v) = \min_{(v,w) \in E} \rho(w) + 1$; and (iv) for $v \in V_2 \setminus B$, we have $\rho(v) = \max_{(v,w) \in E} \rho(w) + 1$. We define the comparison operators $\leq, \geq$ on progress measures with the *pointwise* comparison, i.e., for $\bowtie \in \{\leq, \geq\}$ and progress measures $\rho_1$ and $\rho_2$, we write $\rho_1 \bowtie \rho_2$ iff for all $v \in V$ we have $\rho_1(v) \bowtie \rho_2(v)$.

*Lift operation on progress measure.* Given a game graph $G$, the function $\mathsf{Lift}^G$ takes as input a progress measure and returns a progress measure. For all input progress measures $\rho$, the output progress measure $\rho' = \mathsf{Lift}^G(\rho)$ is defined as follows: for all $v \in V$, (i) for $v \in V_1 \cap B$, we have $\rho'(v) = \top$ if for all $(v, w) \in E$ we have $\rho(w) = \top$, and 0 otherwise; (ii) for $v \in V_2 \cap B$, we have $\rho'(v) = \top$ if there exists $(v, w) \in E$ with $\rho(w) = \top$, and 0 otherwise, (iii) for $v \in V_1 \setminus B$, we have $\rho'(v) = \min_{(v,w) \in E} \rho(w) + 1$; and (iv) for $v \in V_2 \setminus B$, we have $\rho'(v) = \max_{(v,w) \in E} \rho(w) + 1$.

LEMMA 5.1. *For all game graphs $G$, the function $\mathsf{Lift}^G$ is monotonic (if $\rho_1 \leq \rho_2$, then $\mathsf{Lift}^G(\rho_1) \leq \mathsf{Lift}^G(\rho_2)$).*

*Proof.* Consider progress measures $\rho_1, \rho_2$ such that $\rho_1 \leq \rho_2$. For a non-Büchi vertex $v \in (V \setminus B)$ we have

$$\mathsf{Lift}^G(\rho_1)(v) = \begin{cases} \min_{(v,w)\in E} \rho_1(w) + 1 \leq \min_{(v,w)\in E} \rho_2(w) + 1 = \mathsf{Lift}^G(\rho_2)(v) & v \in V_1 \setminus B; \\ \max_{(v,w)\in E} \rho_1(w) + 1 \leq \max_{(v,w)\in E} \rho_2(w) + 1 = \mathsf{Lift}^G(\rho_2)(v) & v \in V_2 \setminus B; \end{cases}$$

where $E$ is the set of edges in $G$. It follows that for all $v \in (V \setminus B)$ we have $\mathsf{Lift}^G(\rho_1)(v) \leq \mathsf{Lift}^G(\rho_2)(v)$. Note that for vertices in $B$, progress measures are either 0 or $\top$. For $v \in B$ we have the following cases: (i) $v \in V_1 \cap B$: if $\mathsf{Lift}^G(\rho_1)(v) = \top$, then for all $(v,w) \in E$ we have $\rho_1(w) = \top$, and hence for all $(v,w) \in E$ we have $\rho_2(w) = \top$; thus $\mathsf{Lift}^G(\rho_2)(v) = \top$; and (i) $v \in V_2 \cap B$: if $\mathsf{Lift}^G(\rho_1)(v) = \top$, then there exists $(v,w) \in E$ with $\rho_1(w) = \top$, and hence we have $\rho_2(w) = \top$; thus $\mathsf{Lift}^G(\rho_2)(v) = \top$. It follows that we have $\mathsf{Lift}^G(\rho_1) \leq \mathsf{Lift}^G(\rho_2)$. The desired result follows. ∎

Since $\mathsf{Lift}^G$ is a monotonic function on a finite lattice, by the Tarski-Knaster Theorem [22] it has a least fix-point. Given a player-1 attractor $Attr_1(U, G)$, the *minimal alternating distance* of a vertex $v \in Attr_1(U, G)$ is the rank $rank(v)$ of the vertex $v$ (in other words it is the alternating shortest distance to $U$ where player-1 minimizes the distance and player-2 maximizes the distance to $U$). The result of [20] established that for all game graphs $G$, (i) there is a unique least fix-point of $\mathsf{Lift}^G$, (ii) the least fix-point is a valid progress measure, (iii) in the winning set the progress measure equals the minimal alternating distance to the set of Büchi vertices in the winning set and all Büchi vertices in the winning set have progress measure 0, and (iv) all vertices in the complement of the winning set are assigned $\top$. The result of [20] is for the more general case of parity objectives, and the specialization to Büchi objectives yields the above properties.

THEOREM 5.1. ([20]) *For all game graphs $G$, let $\rho^*$ be the least fix-point of $\mathsf{Lift}^G$, and let $||\rho^*|| = \{v \in V \mid \rho(v) \in [n]\}$ denote the set of vertices that are not assigned the top element. Then $||\rho^*|| = W_1(\Phi)$, where $\Phi$ is the Büchi objective.*

**Decremental algorithm.** Our algorithm initially computes the least fix-point progress measure $\rho^*$ of the graph and then maintains it after each edge deletion by repeatedly applying the lift operator to the fix-point $\rho^*$ stored *before* the edge deletion. To prove the correctness we will show that the fix-point obtained by repeatedly applying the lift operator on the previous least fix-point converges to the least fix-point of the new game graph. The algorithm maintains the following data structure: (i) For each vertex $x \in V_1 \cap B$ it keeps a list of vertices $w$ such that $(x, w) \in E$ and $\rho^*(w) \neq \top$ and (ii) for each vertex $x \in V_1 \setminus B$ a list of vertices $w$ such that $(x, w) \in E$ and $\rho^*(x) = \rho^*(w) + 1$. (iii) Every edge $(x, w)$ has a pointer to its location in the list of $x$ if it is stored in such a list. We next describe the algorithm in detail.

*Computation of the initial $\rho^*$.* Use the static Büchi algorithm from the previous section to compute the player-1 and player-2 winning sets and assign $\top$ to all vertices in the player-2 winning set. Use the backward search algorithm [2, 19] to determine the rank of every vertex in the player-1 winning set and set its initial progress measure equal to its rank. Then we compute for each vertex of $V_1$ its list.

*Deletion of the edge $(u, v)$.* Maintain a queue of vertices to be processed to update the progress measure until the least fix-point is reached such that a vertex of $V_2$ is only added to the queue when its progress measure has increased. Initially, enqueue $u$. Then iteratively process and dequeue the vertices from the queue.

*Case 1: A vertex $x$ of $V_1$ is dequeued.* Check whether given the current progress measure, the progress measure of $x$ needs to be increased to satisfy the lift operation for $x$. To do this we first check whether the list of $x$ is empty. If it is not empty, nothing needs to be done. If it is empty, all remaining outedges of $x$ are checked to compute the new progress measure value of $x$ and the new list of $x$. Then all inedges $(u, x)$ of $x$ are processed as follows: If $u$ is a player-1 non-Büchi vertex ($u \in V_1 \setminus B$), then it is enqueued (if it is not already in the queue) and $x$ is removed from the list of $u$ if it was there. If $u$ is a player-2 non-Büchi vertex ($u \in V_2 \setminus B$), then check whether the change in the progress measure value of $x$ also increases the progress measure value of $u$. If it does, then $u$ is enqueued (if it is not already in the queue), otherwise $u$ is *not* enqueued. If $u$ is a player-1 Büchi vertex ($u \in V_1 \cap B$), then (i) if the progress measure of $x$ is not $\top$, then do nothing; (ii) else remove $x$ from the list of $u$, and if the list of $u$ is empty, assign progress measure $\top$ to $u$ and $u$ is enqueued (if it is not already in the queue). If $u$ is a player-2 Büchi vertex ($u \in V_2 \cap B$), then (i) if the progress measure of $x$ is not $\top$, then do nothing; (ii) else assign progress measure $\top$ to $u$ and $u$ is enqueued (if it is not already in the queue).

*Case 2: A vertex $x$ of $V_2$ is dequeued.* In this case the progress measure of $x$ has increased and it has already been updated. Thus all what remains is to process all inedges $(u, x)$ of $x$ as follows: If $u$ is a player-1 non-Büchi vertex, then it is enqueued (if it is not already in the queue) and $x$ is removed from the list of $u$ if it was there. If $u$ is a player-2 non-Büchi vertex, then check whether the change in the progress measure value of $x$ also increases the progress measure value of $u$. If it does, then $u$ is enqueued (if it is not already in the queue), otherwise $u$ is *not* enqueued. If $u$ is a player-1 Büchi vertex, then (i) if the progress measure of $x$ is not $\top$, then do nothing; (ii) else remove $x$ from the list of $u$, and if the list of $u$ is empty, assign progress measure $\top$ to $u$ and $u$ is enqueued (if it is not already in the queue). If $u$ is a player-2 Büchi vertex, then (i) if the progress measure of $x$ is not $\top$, then do nothing; (ii) else assign progress measure $\top$ to $u$ and $u$ is enqueued (if it is not already in the queue).

This algorithm is a generalization of the Even-Shiloach algorithm [17] for maintaining the connected component (or more precisely the breadth-first-search tree) of a vertex $b$ in an undirected graph. Assume $B = \{b\}$ and that $V = V_1$. Then the progress measure value of a vertex $v$ is exactly $v$'s level in the breadth-first search tree rooted at $b$ (or equivalently its shortest path distance to $b$). Applying the lift operator to a vertex $v$ is exactly the same as checking whether $v$ has still an edge to an edge at level $level(v) - 1$ and if not, increasing the level of $v$ by 1.

**Correctness.** Let $G$ be a game graph, and let $\rho^*$ be the least fix-point of $\text{Lift}^G$. Let $\overline{G} = G \setminus \{e\}$, where $e \in E \cap V_1 \times V$, be the game graph obtained by deleting a player-1 edge $e$. Let $\overline{\rho}^*$ be the least fix-point of $\overline{G}$. Let $\rho^*_{\text{new}}$ be the new fix-point obtained by iterating $\text{Lift}^{\overline{G}}$ on $\rho^*$. We will show that $\overline{\rho}^* = \rho^*_{\text{new}}$.

LEMMA 5.2. *We have $\overline{\rho}^* \leq \rho^*_{\text{new}}$.*

*Proof.* Let $\rho_0$ be the progress measure that assings 0 to all vertices, i.e., the least progress measure. Clearly, $\rho_0 \leq \rho^*$. Let us denote by $(\text{Lift}^{\overline{G}})^i$ the result of applying the lift operator $i$-times on $\overline{G}$, for some $i \in \mathbb{N}$. From a simple application of Lemma 5.1 it follows that $(\text{Lift}^{\overline{G}})^i$ is monotonic. Hence we have $(\text{Lift}^{\overline{G}})^i(\rho_0) \leq (\text{Lift}^{\overline{G}})^i(\rho^*)$. Since $\overline{\rho}^* = (\text{Lift}^{\overline{G}})^j(\rho_0)$ for some $j$, and $\rho^*_{\text{new}} \geq (\text{Lift}^{\overline{G}})^i(\rho^*)$ for all $i$ (in particular for the $j$ for which the least fix-point is obtained from $\rho_0$), it follows that $\overline{\rho}^* \leq \rho^*_{\text{new}}$. ∎

LEMMA 5.3. *We have $\rho^*_{\text{new}} \leq \overline{\rho}^*$.*

*Proof.* Observe that the graph $\overline{G}$ is obtained by deleting an edge for player-1, and hence the winning set for player 1 can only decrease and the minimal alternating distance to the Büchi set in the winning set can only increase. In other words, we have $\rho^* \leq \overline{\rho}^*$, i.e., the least fix-point of the graph $G$ is smaller than the least fix-point of $\overline{G}$. Since $\rho^*_{\text{new}} = (\text{Lift}^{\overline{G}})^i(\rho^*)$, for some $i$, we have $\rho^*_{\text{new}} = (\text{Lift}^{\overline{G}})^i(\rho^*) \leq (\text{Lift}^{\overline{G}})^i(\overline{\rho}^*) = \overline{\rho}^*$, where the first inequality is a consequence of Lemma 5.1 that $(\text{Lift}^{\overline{G}})^i$ is monotonic, and the last inequality is a consequence of the fact that $\overline{\rho}^*$ is a fix-point. Hence the desired result follows. ∎

The correctness follows from Lemma 5.2 and Lemma 5.3 (that $\rho^*_{\text{new}} = \overline{\rho}^*$) and the fact that the algorithm implements the iteration of the lift operator on vertices one by one to compute the fix-point that is obtained by repeatedly applying the lift operator on the least fix-point of the previous game graph.

**Running time.** The deletions of player-1 edges only decreases the winning set, and once a set is removed from the winning set (i.e., assigned value $\top$ in the progress measure algorithm), then they are never worked upon. Upon termination, let $W$ be the winning set, and let $\rho$ be the least fix-point in the end. The computation of the initial least fix-point is done in time $O(n^2)$.

In the decremental algorithm we check for each dequeued player-1 vertex $u$ whether its progress measure increases in constant time. If it does not increase no further work is done for $u$. The constant amount of work is charged to the edge deletion if an outedge of $u$ was deleted. If no outedge of $u$ was deleted then the progress measure of a vertex $w$ with $(u, w) \in E$ must have increased and we charge the work to $w$. If the progress measure of $u$ increases we spend time $O(|\mathsf{In}(u)| + |\mathsf{Out}(u)|)$ to determine the new progress measure of $u$, compute its new list, and process all its inedges, and the work is charged to $u$. A player-2 vertex $u$ is only enqueued when its progress measure has increased. When it is dequeued we spend time $O(|\mathsf{In}(u)|)$ to process all its inedges, and charge it to $u$. The number of times the progress measure can increase for a vertex is at most $n + 1$ (as once it is $n + 1$ it is assigned $\top$). For a vertex $v$, let $\mathsf{Num}(v) = \rho(v)$, if $\rho(v) \neq \top$, and $n + 1$ otherwise. Hence the total work done by the algorithm is

$$O(\sum_{v \in V} \mathsf{Num}(v) \cdot |\mathsf{In}(v)|) + O(\sum_{v \in V} \mathsf{Num}(v) \cdot |\mathsf{Out}(v)|) = O(n \cdot m).$$

THEOREM 5.2. *Given an initial game graph with $n$ vertices and $m$ edges, the winning set partitions can be maintained under the deletion of $O(m)$ edges $(u, v)$ with $u \in V_1$ in total time $O(n \cdot m)$.*

**5.2 Incremental algorithm for Büchi games** We now present the details of the incremental algorithm for Büchi games, where we consider insertion of player-1 edges. The algorithm is almost identical to the decremental algorithm and based on the dual progress measure for player 2. We start with the definition of a valid progress measure for player 2.

*Valid progress measure for player 2.* Given a game graph with a set $B$ of Büchi vertices, let $C = V \setminus B$ be the set of coBüchi vertices. A progress measure $\rho$ is a *valid* progress measure for player 2 if the following conditions hold for all $v \in V$:

$$\rho(v) \geq \begin{cases} \min_{(v,w) \in E} \rho(w) & v \in V_2 \cap C; \\ \min_{(v,w) \in E} \rho(w) + 1 & v \in V_2 \cap B; \\ \max_{(v,w) \in E} \rho(w) & v \in V_1 \cap C; \\ \max_{(v,w) \in E} \rho(w) + 1 & v \in V_1 \cap B. \end{cases}$$

We define the comparison operators $\leq, \geq$ on progress measures with the *pointwise* comparison.

*Lift operation on progress measure.* Given a game graph $G$, the function $\mathsf{coLift}^G$, like the $\mathsf{Lift}^G$ function, takes as input a progress measure and returns a progress measure. For all input progress measures $\rho$, the output progress measure $\rho' = \mathsf{coLift}^G(\rho)$ is defined as follows: for all $v \in V$,

$$\rho'(v) = \begin{cases} \min_{(v,w) \in E} \rho(w) & v \in V_2 \cap C; \\ \min_{(v,w) \in E} \rho(w) + 1 & v \in V_2 \cap B; \\ \max_{(v,w) \in E} \rho(w) & v \in V_1 \cap C; \\ \max_{(v,w) \in E} \rho(w) + 1 & v \in V_1 \cap B. \end{cases}$$

LEMMA 5.4. *For all game graphs $G$, the function $\mathsf{coLift}^G$ is monotonic.*

*Proof.* Consider progress measures $\rho_1, \rho_2$ such that $\rho_1 \leq \rho_2$. For a vertex $v$ we have

$$\mathsf{coLift}^G(\rho_1)(v) = \begin{cases} \min_{(v,w) \in E} \rho_1(w) \leq \min_{(v,w) \in E} \rho_2(w) = \mathsf{coLift}^G(\rho_2)(v) & v \in V_2 \cap C; \\ \min_{(v,w) \in E} \rho_1(w) + 1 \leq \min_{(v,w) \in E} \rho_2(w) + 1 = \mathsf{coLift}^G(\rho_2)(v) & v \in V_2 \cap B; \\ \max_{(v,w) \in E} \rho_1(w) \leq \max_{(v,w) \in E} \rho_2(w) = \mathsf{coLift}^G(\rho_2)(v) & v \in V_1 \cap C; \\ \max_{(v,w) \in E} \rho_1(w) + 1 \leq \max_{(v,w) \in E} \rho_2(w) + 1 = \mathsf{coLift}^G(\rho_2)(v) & v \in V_1 \cap B; \end{cases}$$

where $E$ is the set of edges in $G$. It follows that $\mathsf{coLift}^G(\rho_1) \leq \mathsf{coLift}^G(\rho_2)$. The desired result follows. ∎

Since coLift$^G$ is a monotonic function on a finite lattice, by Tarski-Knaster Theorem [22] it has a least fix-point. Before we proceed to the characterization, we present a definition: for a vertex $v \in W_2(\Psi)$, where $\Psi$ is the coBüchi objective coBuchi($C$), let maxvisit($v$) = $\min_{\pi \in \Pi} \max_{\sigma \in \Sigma} |\{i \mid \omega(v, \sigma, \pi) = \langle v_0, v_1, v_2, \ldots \rangle, v_i \in B\}|$ denote the maximum number of visits to Büchi vertices. Since $v \in W_2(\Psi)$, once a winning strategy for player-2 is fixed, there cannot be a cycle with a Büchi vertex, and hence maxvisit($v$) $\leq n$. The result of [20] established that for all game graphs $G$, (i) there is a unique least fix-point of coLift$^G$, (ii) the least fix-point is a valid progress measure, (iii) for vertices $v$ in the winning set for player 2 the progress measure equals maxvisit($v$), and (iv) all vertices in the winning set for player 1 are assigned the top element $\top$. The result of [20] is for the more general case of parity objectives, and the specialization to coBüchi objectives yields the above properties.

THEOREM 5.3. ([20]) *For all game graphs G, let $\rho^*$ be the least fix-point of coLift$^G$, and let $||\rho^*|| = \{v \in V \mid \rho(v) \in [n]\}$ denote the set of vertices that are not assigned the top element. Then $||\rho^*|| = W_2(\Psi)$, where $\Psi$ is the coBüchi objective.*

**Incremental algorithm.** Our algorithm initially computes the least fix-point progress measure $\rho^*$ of coLift of the graph and then maintains it after each edge insertion by repeatedly applying the lift operator coLift to the fix-point $\rho^*$ stored from *before* the edge insertion. To prove the correctness we will show that the fix-point obtained by repeatedly applying the lift operator on the previous least fix-point converges to the least fix-point of the new game graph. The algorithm maintains the following data structure: (i) For each vertex $x \in V_2 \cap C$ it keeps a list of vertices $w$ such that $(x, w) \in E$ and $\rho^*(x) = \rho^*(w)$ and (ii) for each vertex $x \in V_2 \cap B$ a list of vertices $w$ such that $(x, w) \in E$ and $\rho^*(x) = \rho^*(w) + 1$. (iii) Every edge $(x, w)$ has a pointer to its location in the list of $x$ if it is stored in such a list. We next describe the algorithm in detail. We first describe the insertion of an edge as the initial fix-point computation is similar.

*Insertion of the edge $(u, v)$.* Maintain a queue of vertices to be processed to update the progress measure until the least fix-point is reached such that a vertex of $V_1$ is only added to the queue when its progress measure has increased. Initially, enqueue $u$. Then iteratively process and dequeue the vertices from the queue.

  *Case 1: A vertex $x$ of $V_2$ is dequeued.* Check whether given the current progress measure, the progress measure of $x$ needs to be increased to satisfy the lift operation for $x$. To do this we first check whether the list of $x$ is empty. If it is not empty, nothing needs to be done. If it is empty, all remaining outedges of $x$ are checked to compute the new progress measure value of $x$ and the new list of $x$. Then all inedges $(u, x)$ of $x$ are processed as follows: If $u$ is a player-2 vertex it is enqueued (if it is not already in the queue) and $x$ is removed from the list of $u$ if it was there. If $u$ is a player-1 vertex then check whether the change in the progress measure value of $x$ also increases the progress measure value of $u$. If it does, then $u$ is enqueued (if it is not already in the queue), otherwise $u$ is *not* enqueued.

  *Case 2: A vertex $x$ of $V_1$ is dequeued.* In this case the progress measure of $x$ has increased and it has already been updated. Thus all what remains is to process all inedges $(u, x)$ of $x$ as follows: If $u$ is a player-2 vertex it is enqueued (if it is not already in the queue) and $x$ is removed from the list of $u$ if it was there. If $u$ is a player-1 vertex then check whether the change in the progress measure value of $x$ also increases the progress measure value of $u$. If it does, then $u$ is enqueued (if it is not already in the queue), otherwise $u$ is *not* enqueued.

*Computation of the initial $\rho^*$.* The computation of the initial $\rho^*$ is similar to the incremental algorithm. We initialize the initial progress measure as 0 for all vertices, then enqueue the set of Büchi vertices, and proceed as the incremental algorithm until a fix-point is reached. As we start with the all 0 progress measure and repeatedly apply the lift operator we are guaranteed to reach the least fix-point. Then we compute for each vertex $v \in V_2$ its list.

*Correctness.* Let $G$ be a game graph, and let $\rho^*$ be the least fix-point of coLift$^G$. Let $\overline{G} = G \cup \{e\}$, where $e \in E \cap V_1 \times V$, be the game graph obtained by inserting a player-1 edge $e$. Let $\overline{\rho}^*$ be the least fix-point of $\overline{G}$. Let $\rho^*_{\text{new}}$ be the new fix-point obtained by iterating coLift$^{\overline{G}}$ on $\rho^*$. We will show that $\overline{\rho}^* = \rho^*_{\text{new}}$.

LEMMA 5.5. *We have $\overline{\rho}^* \leq \rho_{\mathsf{new}}^*$.*

*Proof.* Let $\rho_0$ be the progress measure that assings 0 to all vertices, i.e., the least progress measure. Clearly, $\rho_0 \leq \rho^*$. Let us denote by $(\mathsf{coLift}^{\overline{G}})^i$ the result of applying the lift operator $i$-times on $\overline{G}$, for some $i \in \mathbb{N}$. From a simple application of Lemma 5.4 it follows that $(\mathsf{coLift}^{\overline{G}})^i$ is monotonic. Hence we have $(\mathsf{coLift}^{\overline{G}})^i(\rho_0) \leq (\mathsf{coLift}^{\overline{G}})^i(\rho^*)$. Since $\overline{\rho}^* = (\mathsf{coLift}^{\overline{G}})^j(\rho_0)$ for some $j$, and $\rho_{\mathsf{new}}^* \geq (\mathsf{coLift}^{\overline{G}})^i(\rho^*)$ for all $i$ (in particular for the $j$ for which the least fix-point is obtained from $\rho_0$), it follows that $\overline{\rho}^* \leq \rho_{\mathsf{new}}^*$. ∎

LEMMA 5.6. *We have $\rho_{\mathsf{new}}^* \leq \overline{\rho}^*$.*

*Proof.* Observe that the graph $\overline{G}$ is obtained by inserting an edge for player-1, and hence the winning set for player 2 can only decrease and $\mathsf{maxvisit}(v)$ can only increase for vertices in the winning set for player 2. In other words, we have $\rho^* \leq \overline{\rho}^*$, i.e., the least fix-point of the graph $G$ is smaller than the least fix-point of $\overline{G}$. Since $\rho_{\mathsf{new}}^* = (\mathsf{coLift}^{\overline{G}})^i(\rho^*)$, for some $i$, we have

$$\rho_{\mathsf{new}}^* = (\mathsf{coLift}^{\overline{G}})^i(\rho^*) \leq (\mathsf{coLift}^{\overline{G}})^i(\overline{\rho}^*) = \overline{\rho}^*,$$

where the first inequality is a consequence of Lemma 5.4 that $(\mathsf{coLift}^{\overline{G}})^i$ is monotonic, and the last inequality is a consequence of the fact that $\overline{\rho}^*$ is a fix-point. Hence the desired result follows. ∎

LEMMA 5.7. *We have $\rho_{\mathsf{new}}^* = \overline{\rho}^*$.*

**Correctness.** The correctness follows from Lemma 5.7 and the fact that the algorithm implements the iteration of the lift operator on vertices one by one to compute the fix-point that is obtained by repeatedly applying the lift operator on the least fix-point of the previous game graph.

**Running time.** The insertions of player-1 edges only decreases the winning set for player 2, and once a set is removed from the winning set (i.e., assigned value $\top$ in the progress measure algorithm), then they are never worked upon. Upon termination, let $W$ be the winning set, and let $\rho$ be the least fix-point in the end. In the incremental algorithm we check for each dequeued player-2 vertex $u$ whether its progress measure increases in constant time. If it does not increase no further work is done for $u$. Since $u$ is processed, the progress measure of a vertex $w$ with $(u, w) \in E$ must have increased and we charge the work to $w$. If the progress measure of $u$ increases, then we spend time $O(|\mathsf{In}(u)| + |\mathsf{Out}(u)|)$ to determine the new progress measure of $u$, compute its new list, and process all its inedges, and charge the work to $u$. A player-1 vertex $u$ is only enqueued when its progress measure has increased, or an edge is inserted at $u$. If an edge was inserted, the work is charged to the inserted edge. When it is dequeued we spend time $O(|\mathsf{In}(u)|)$ to process all its inedges, and charge it to $u$. The number of times the progress measure can increase for a vertex is at most $n + 1$ (as once it is $n + 1$ it is assigned $\top$). For a vertex $v$, let $\mathsf{Num}(v) = \rho(v)$, if $\rho(v) \neq \top$, and $n + 1$ otherwise. Hence the total work done by the algorithm is

$$O(\sum_{v \in V} \mathsf{Num}(v) \cdot |\mathsf{In}(v)|) + O(\sum_{v \in V} \mathsf{Num}(v) \cdot |\mathsf{Out}(v)|) = O(n \cdot m).$$

An argument similar to the above also establishes that the initial least fix-point is computed in time $O(n \cdot m)$.

THEOREM 5.4. *Given an initial game graph with $n$ vertices and $m$ edges, the winning set partitions can be maintained under the insertion of $O(m)$ edges $(u, v)$ with $u \in V_1$ in total time $O(n \cdot m)$.*

**Acknowledgements.** The research was supported by Austrian Science Fund (FWF) Grant No P 23499-N23 on Modern Graph Algorithmic Techniques in Formal Verification, ERC Start grant (279307: Graph Games), and Microsoft faculty fellows award.

# References

[1] R. Alur, T.A. Henzinger, and O. Kupferman. Alternating-time temporal logic. *Journal of the ACM*, 49:672–713, 2002.

[2] C. Beeri. On the membership problem for functional and multivalued dependencies in relational databases. *ACM Trans. on Database Systems*, 5:241–259, 1980.

[3] J.R. Büchi. Weak second-order arithmetic and finite automata. *Zeitschrift für mathematische Logik und Grundlagen der Mathematik*, 6:66–92, 1960.

[4] J.R. Büchi. On a decision method in restricted second-order arithmetic. In E. Nagel, P. Suppes, and A. Tarski, editors, *Proceedings of the First International Congress on Logic, Methodology, and Philosophy of Science 1960*, pages 1–11. Stanford University Press, 1962.

[5] J.R. Büchi and L.H. Landweber. Solving sequential conditions by finite-state strategies. *Transactions of the AMS*, 138:295–311, 1969.

[6] A. K. Chandra, D. Kozen, and L. J. Stockmeyer. Alternation. *J. ACM*, 28(1):114–133, 1981.

[7] K. Chatterjee and M. Henzinger. Faster and dynamic algorithms for maximal end-component decomposition and related graph problems in probabilistic verification. In *SODA'11*. SIAM, 2011.

[8] K. Chatterjee, T.A. Henzinger, and N. Piterman. Algorithms for Büchi games. In *Games in Design and Verification (GDV)*, 2006.

[9] K. Chatterjee, M. Jurdziński, and T.A. Henzinger. Simple stochastic parity games. In *CSL'03*, volume 2803 of *LNCS*, pages 100–113. Springer, 2003.

[10] K. Chatterjee, M. Jurdziński, and T.A. Henzinger. Quantitative stochastic parity games. In *SODA'04*, pages 121–130. SIAM, 2004.

[11] A. Church. Logic, arithmetic, and automata. In *Proceedings of the International Congress of Mathematicians*, pages 23–35. Institut Mittag-Leffler, 1962.

[12] A. Condon. The complexity of stochastic games. *Information and Computation*, 96(2):203–224, 1992.

[13] C. Courcoubetis and M. Yannakakis. The complexity of probabilistic verification. *Journal of the ACM*, 42(4):857–907, 1995.

[14] L. de Alfaro and T.A. Henzinger. Interface automata. In *FSE'01*, pages 109–120. ACM Press, 2001.

[15] D.L. Dill. *Trace Theory for Automatic Hierarchical Verification of Speed-independent Circuits*. The MIT Press, 1989.

[16] E.A. Emerson and C. Jutla. Tree automata, mu-calculus and determinacy. In *FOCS'91*, pages 368–377. IEEE, 1991.

[17] S. Even and Y. Shiloach. An on-line edge-deletion problem. *J. ACM*, 28(1):1–4, 1981.

[18] M. R. Henzinger, V. King, and T. Warnow. Constructing a tree from homeomorphic subtrees, with applications to computational evolutionary biology. *Algorithmica*, 24(1):1–13, 1999.

[19] N. Immerman. Number of quantifiers is better than number of tape cells. *Journal of Computer and System Sciences*, 22:384–406, 1981.

[20] M. Jurdziński. Small progress measures for solving parity games. In *STACS'00*, pages 290–301. LNCS 1770, Springer, 2000.

[21] M. Jurdziński, O. Kupferman, and T. A. Henzinger. Trading probability for fairness. In *CSL: Computer Science Logic*, Lecture Notes in Computer Science 2471, pages 292–305. Springer, 2002.

[22] A. Kechris. *Classical Descriptive Set Theory*. Springer, 1995.

[23] O. Kupferman and M.Y. Vardi. From linear time to branching time. *ACM Transactions on Computational Logic*, 6(2):273–294, 2005.

[24] A. Mahanti and A. Bagchi. AND/OR graph heuristic search methods. *JACM*, 32(1):28–51, 1985.

[25] R. McNaughton. Infinite games played on finite graphs. *Annals of Pure and Applied Logic*, 65:149–184, 1993.

[26] A. Pnueli and R. Rosner. On the synthesis of a reactive module. In *POPL'89*, pages 179–190. ACM Press, 1989.

[27] P.J. Ramadge and W.M. Wonham. Supervisory control of a class of discrete-event processes. *SIAM Journal of Control and Optimization*, 25(1):206–230, 1987.

[28] W. Thomas. Languages, automata, and logic. In G. Rozenberg and A. Salomaa, editors, *Handbook of Formal Languages*, volume 3, Beyond Words, chapter 7, pages 389–455. Springer, 1997.

[29] M.Y. Vardi. Automata-theoretic model checking revisited. In *Proc. of Verification, Model Checking, and Abstract Interpretation*, volume LNCS 4349, pages 137–150. Springer, 2007.

[30] M.Y. Vardi. The Büchi complementation saga. In *Proc. of Symp. on Theoretical Aspects of Computer Science*, volume LNCS 4393, pages 12–22. Springer, 2007.

[31] W. Zielonka. Infinite games on finitely coloured graphs with applications to automata on infinite trees. In *Theoretical Computer Science*, volume 200(1-2), pages 135–183, 1998.